\documentclass[
	onecolumn,
	superscriptaddress,
	aps,
	pre,
	amsmath,
	amssymb,
	floatfix,
	nofootinbib,
	notitlepage
]{revtex4-2}

\usepackage{amsmath}
\usepackage{amsfonts}
\usepackage{amssymb}
\usepackage{empheq}
\usepackage{graphicx}
\usepackage{longtable}
\usepackage{dcolumn}
\usepackage{bm}
\usepackage{epstopdf}
\usepackage{datetime}
\usepackage{comment}
\usepackage{empheq}
\usepackage{mathtools}
\usepackage{amssymb}
\longdate

\def\eq#1{{\ref{#1}}}

\usepackage{color}

\begin{document}

\title{Tuning nonlinear waves in nonreciprocal active filaments}
 
\author{Sami C.~Al-Izzi}
\thanks{These authors contributed equally}
\email{\texttt{s.al-izzi@unsw.edu.au}}
\affiliation{Department of Applied Mathematics, School of Mathematics and Statistics, UNSW, Sydney, NSW 2052, Australia.}

\author{Jack Binysh}
\thanks{These authors contributed equally}
\email{\texttt{j.a.c.binysh@uva.nl}}
\affiliation{Department of Mechanical Engineering, University of Birmingham, Birmingham, B15 2TT, UK}
\affiliation{Institute of Physics, Universiteit van Amsterdam, Science Park 904, 1098 XH Amsterdam, The Netherlands}

\author{Yao Du}
\affiliation{Institute of Physics, Universiteit van Amsterdam, Science Park 904, 1098 XH Amsterdam, The Netherlands}

\author{Corentin Coulais}
\affiliation{Institute of Physics, Universiteit van Amsterdam, Science Park 904, 1098 XH Amsterdam, The Netherlands}

\author{Andreas Carlson}\email{\texttt{acarlson@math.uio.no}}
\affiliation{Department of Mathematics, Faculty of Mathematics and Natural Sciences, University of Oslo, 0315 Oslo, Norway}
\affiliation{Ume{\aa} University, Department of Medical Biochemistry and Biophysics, 901 87 Ume{\aa}, Sweden}

\begin{abstract}
The instabilities of slender structures power biological locomotion across scales, and offer a compelling method to actuate soft robots. Nonreciprocal elastic solids have been found to amplify flexural waves in one direction only, but design principles to tune and stabilize these waves are missing. Here we develop a geometrically exact theory of nonreciprocal filaments and provide simulations that capture their post-instability nonlinear dynamics. We find that nonreciprocity, when coupled to inertia or pre-stress, amplifies and advects curvature variations. The resulting one-way patterns of shape morphing can then be selected via dissipative interactions with the environment. Our work offers a continuum-based strategy for how internal stresses can drive active unidirectional waves without need for additional degrees of freedom.
\end{abstract}

\maketitle
\section{Introduction}
Active filaments are a workhorse for mechanical actuation. Driven slender structures pump fluid~\cite{cammannFormFunctionBiological2025a}, squeeze through constrictions~\cite{xiEmergentBehaviorsBucklingdriven2024}, navigate mazes~\cite{wangMechanicalIntelligenceSimplifies2023} and gather materials~\cite{sinaasappelParticleSweepingCollection2026}. This repertoire of functions stems from two ingredients: geometrical slenderness to enable bending and in turn large deformations, and directed waves to drive cyclic motion. A central challenge spanning biophysics~\cite{cammannFormFunctionBiological2025a} and soft robotics \cite{laschiSoftRoboticsTechnologies2016,dengNonlinearWavesFlexible2021, vanlaakeBioinspiredAutonomySoft2024} is to abstract minimal mechanisms for generating such persistent bending waves.

Continuum theories of filament dynamics solve force balance equations for the centerline ${\bf x}$ of the filament, generically written as $\rho\ddot{\bf x}+\beta\dot{\bf x}=\nabla\cdot\boldsymbol{\sigma}+{\bf f}$,
where $\boldsymbol{\sigma}$ is the internal stress in the filament, ${\bf f}$ is the applied force, and $\rho$ and $\beta$ are the line density and ambient friction. To actuate flexural waves these theories point to two physical mechanisms: apply external forces ${\bf f}$, or orchestrate internal stresses ${\boldsymbol{\sigma}}$.  

The first mechanism includes boundary driven forcing~\cite{goldstein1998,wolgemuth2000}, and the follower force chain or active Brownian polymer~\cite{sekimoto1995, chelakkotFlagellarDynamicsConnected2014, decanio2017, fily2020,clarke2024,winklerPhysicsActivePolymers2020}. In follower force chains, internal stresses are purely passive, but a force density ${\bf f}\sim{\bf t}$ is applied to the filament along the tangent ${\bf t}$. Follower force models effectively capture the snapping of natural filaments and guide the design of artificial structures~\cite{zheng2023, martinetEmergentDynamicsActive2025a, weiAutonomousLifelikeBehavior2025}. However, these models rely crucially on external forces, not internally driven stresses---without a substrate to push off, they cannot generate oscillations. This lack of internal dynamics presents a challenge in modelling natural beating patterns driven by net-force-free processes, and restricts the possibilities of follower-force chains in designing artificial locomotors.  

A second mechanism orchestrates stress waves using internal variables that often represent chemical concentration fields or neurological activity. For example, reaction diffusion systems are proposed to explain the long-wavelength beating of eukaryotic cilia~\cite{cassReactiondiffusionBasisAnimated2023} and central pattern generators---with their continuum description as phase diffusion equations---capture the locomotion of nematodes and snakes \cite{hu2009,guo2008,ishimoto_robust_2025}. These theories produce persistent oscillations without external drive, but at the cost of complexity: living locomotors organize undulatory waves through layers of mechano-chemical feedback. Here we ask: How can we capture persistent nonlinear oscillations without the need for external forces or additional internal variables? 

In this context, a distinct approach to unidirectional wave amplification is to break the reciprocal symmetries of elastic interactions directly [Fig.~\ref{fig:Phenomenology}(a,b)] \cite{scheibner2020}. Nonreciprocal chains have been shown to boost linear waves \cite{veenstra2025,al-izzi2025}, host one-way solitons \cite{veenstraNonreciprocalTopologicalSolitons2024}, and persistently oscillate when buckled \cite{al-izzi2025}. These phenomena offer a toolkit of beating patterns and locomotion gaits that break time reversal symmetry, are net force-and-torque free, and do not require additional variables.
However, beyond recent work on micropolar chains of hydrodynamic squirmers \cite{nemeth2024,nemeth2025,yan2025,warda2025,nemethShapespaceDynamicsGeometric2026}, the large amplitude bending modes of these active filaments remain unexplored. In particular, a foundational unanswered question is how activity conspires with key physical parameters---inertia, dissipation, external stresses---to select nonlinear bending waves. 

In this work we develop a symmetry-based continuum theory of nonreciprocal, inextensible, unshearable filaments which is geometrically exact and captures large deformations. A linear analysis shows that these filaments are unstable to a long-wavelength oscillatory instability, driven by undirectional amplification of curvature variations by nonreciprocity. The dominant wavenumber of this instability can then be tuned by internal dissipation or an external substrate. Even below the threshold for such spontaneous oscillations, a second oscillatory instability emerges when activity is combined with pre-stress, for example by pointwise external compression or buckling.  We then develop numerical solutions which show that curvature amplification persists into the nonlinear regime, resulting in large-scale, persistent cycles of shape morphing that are robust to perturbation. 

\section{Continuum Theory}
\begin{figure}
    \centering
    \includegraphics[width=\textwidth]{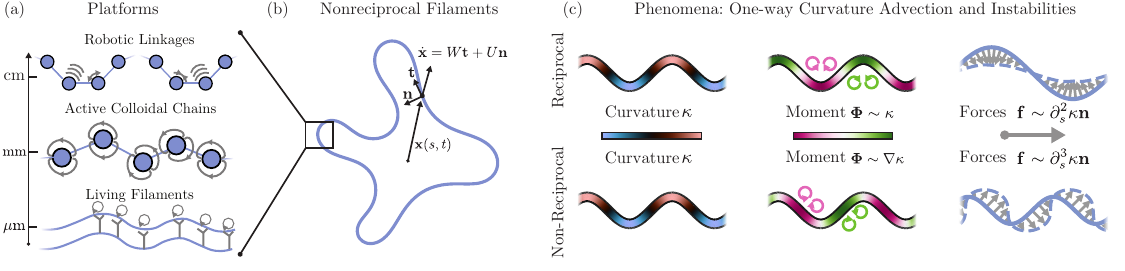}
    \caption{\textbf{Nonreciprocal continuum filaments} 
    (a) Platforms for nonreciprocal bending waves across scales: From living filaments and driven colloidal chains on the microscale~\cite{beatusPhononsOnedimensionalMicrofluidic2006, chajwaWavesAlgebraicGrowth2020, nemeth2025}, to robotic metamaterials with reciprocity-breaking actuation schemes on the macroscale~\cite{al-izzi2025}. 
    (b,c) We consider the continuum limit of such platforms: an active filament whose bending moments break reciprocal symmetry. The filament ${\bf x}(s,t)$ is described by a tangent ${\bf t}$ and normal ${\bf n}$, with velocity $\dot{{\bf x}}=W{\bf t}+U {\bf n}$. Under reciprocal bending elasticity, variations in curvature $\kappa$ lead to moments $\boldsymbol{\Phi}\sim \kappa$ and forces ${\bf f} \sim \partial^2_s \kappa \bf n$, which flatten undulations. By contrast, nonreciprocal interactions coarse-grain to bending moments $\boldsymbol{\Phi}\sim \nabla \kappa$ and forces ${\bf f} \sim \partial^3_s \kappa \bf n$, which advect undulations without attenuation.}
    \label{fig:Phenomenology}
\end{figure}

We first develop a generic continuum description of an inextensible, unshearable rod driven by internal stresses that break reciprocity. We approach this problem in a geometric manner, in analogy to the filament dynamics described in \cite{goldstein1995,goldstein1998,goldstein2000,goldstein2006}. We take a planar filament, described by a centreline ${\bf x} (s,t)$ and Frenet frame $\{{\bf t}(s,t), {\bf n}(s,t), {\bf b}(s,t)\}$, which vary with arclength $s$ and time $t$ [Fig.~\ref{fig:Phenomenology}(b)] and ${\bf b}={\bf e}_z$ is the binormal to the plane. Force balance reads 
\begin{equation}
    \rho\ddot{\bf x}+\beta\dot{\bf x}=\nabla\cdot\boldsymbol{\sigma}
    \label{eq:ForceBalance2}
\end{equation}
where $\nabla={\bf t}\partial_s$. We focus on the case of no external forcing, and will consider three separate contributions to the total stress tensor $\boldsymbol{\sigma}$: 
\begin{equation}
{\boldsymbol \sigma}={\boldsymbol \sigma}^\text{el}+ {\boldsymbol \sigma}^\text{visc}+{\boldsymbol \sigma}^\text{nr}.
\end{equation}
The crucial ingredient will be ${\boldsymbol \sigma}^\text{nr}$, a reciprocity-breaking stress that cannot be derived from a free energy, complemented by a 
passive bending elasticity ${\boldsymbol \sigma}^\text{el}$ and viscous dissipation ${\boldsymbol \sigma}^\text{visc}$ within the filament itself~\cite{kodio_thesis}. 

In order for the stress to conserve angular momentum the antisymmetric part must be derivable from a bending moment tensor, 
${\boldsymbol \Phi}$, has the form
\begin{equation}
    {\boldsymbol \Phi}= \Phi_{tb}{\bf t}{\bf b}\text{.}
\end{equation}
The stress and bending moment are related by $\boldsymbol{\epsilon}:\boldsymbol{\sigma}=\nabla\cdot\boldsymbol{\Phi}$, where $\boldsymbol{\epsilon}$ is the rank-$3$ totally antisymmetric tensor, which in components of the frame field reads
\begin{equation}
    \sigma_{tn} = \partial_s\Phi_{tb}\text{.}
\end{equation}
To recover the more usual bending moment vectors used in Kirchhoff or Cosserat rod theory, ${\bf M}$, one can simply use the relation ${\bf M}={\bf t}\cdot\boldsymbol{\Phi}$. We now construct expressions for $\boldsymbol{\sigma}$ and $\boldsymbol{\Phi}$ directly in terms of the filament geometry.

\subsection{Nonreciprocal stresses versus reciprocal bending elasticity} 
The classical bending moment, $\boldsymbol{\Phi}^{\mathrm{el}}$ for a Kirchoff rod is given by~\cite{audolyElasticityGeometryHair2018}
\begin{equation}
\boldsymbol{\Phi}^{\mathrm{el}}=A (\kappa-\kappa_0) {\bf t} {\bf b},
   \label{eq:phiel}
\end{equation}
with $A$ the bending rigidity, $\kappa$ the geodesic curvature, and $\kappa_0$, the spontaneous curvature. The associated stress tensor, ${\boldsymbol \sigma}^\text{el}$, is given by
\begin{equation}
    {\boldsymbol \sigma}^\text{el} =  \Lambda {\bf t} {\bf t} + A \partial_s\kappa {\bf t}{\bf n} 
    - \frac{A}{2}\left(\kappa^2-\kappa_0^2\right){\bf t}{\bf t}\text{,}
   \label{eq:sigel}
\end{equation}
with symmetric parts chosen to correspond to the variation of the classical energy of a rod with line tension $\Lambda$~\cite{goldstein1995}. Taking the divergence of~\eqref{eq:sigel} leads to the classical forces on a rod,
\begin{equation}
    {\bf f} = \nabla\cdot{\boldsymbol \sigma}^{\mathrm{el}} = \partial_s \Lambda {\bf t} - \kappa \Lambda {\bf n} 
    + A \partial_s^2 \kappa{\bf n}
    + \frac{1}{2 }A \kappa\left(\kappa^2-\kappa_0^2\right){\bf n}\text{.}
   \label{eq:fel}
\end{equation}

Each of these expressions is even under a reversal of the polarity of the filament: $s \rightarrow -s $, $\bf t \rightarrow -\bf t$, $\bf n \rightarrow -\bf n $, $\kappa \rightarrow -\kappa$. 

To break reciprocal symmetry, we now include an active component to the elastic bending moment that is isotropic but explicitly breaks this polar symmetry. The simplest such bending moment is 
\begin{equation}
\boldsymbol{\Phi}^\mathrm{nr}=\zeta \nabla\kappa {\bf b}\text{,}    
    \label{eq:phiodd}
\end{equation} 
with associated stress
\begin{equation}
    {\boldsymbol \sigma}^\text{nr}= \zeta \partial_s^{2}\kappa {\bf t}{\bf n}\text{,}
    \label{eq:signr}
\end{equation}
where $\zeta$ is the nonreciprocity. Taking the divergence of this stress gives 
\begin{equation}
    {\bf f} = \nabla\cdot{\boldsymbol \sigma^{\mathrm{nr}}}  = \zeta \partial_s^{3} \kappa {\bf n} + \zeta \kappa \partial_s^{2} \kappa {\bf t}\text{.}
    \label{eq:fodd}
\end{equation}
Intuitively, the nonreciprocal moment ${\boldsymbol \Phi}^\text{nr}$ and stress ${\boldsymbol \sigma}^\text{nr}$ are related to their elastic counterparts ${\boldsymbol \Phi}^\text{el}$,  ${\boldsymbol \sigma}^\text{el}$ by a single additional arclength gradient. For example, the conventional Kirchoff bending moment~\eqref{eq:phiel} couples directly to curvature, whilst
the nonreciprocal moment~\eqref{eq:phiodd} couples to gradients in curvature. Lower order active stresses, for example ${\boldsymbol \sigma}^\text{nr} \sim \zeta \kappa {\bf t}{\bf n}$, are possible if the assumption of unshearability is relaxed or micropolar ordering is allowed, as is the case in a more general Cosserat rod---such active terms have recently been studied in the context of filaments made of active colloids~\cite{nemeth2025,yan2025,nemeth_shape-space_2026}. 
In contrast to conventional bending elasticity, which flattens curvature variations, these nonreciprocal couplings advect gradients of curvature without attenuation. For example, given a sinusoidal filament of wavenumber $q$, ${\bf x}(s)= (s, \sin{q s})$, the conventional bending moment $\boldsymbol{\Phi}^{\mathrm{el}} \sim q^2 \sin{q s}$ has maxima at the peaks of the sine wave, flattening the filament
[Fig.~\ref{fig:Phenomenology}(c)]. In contrast, $\boldsymbol{\Phi}^{\mathrm{nr}} \sim q^3 \cos{q s}$: the maximal torque dipole lags $\pi/2$ behind the filament geometry, peaking at the inflection points of the curve and driving wave propagation.

\subsection{Viscous dissipation}
The odd stress~\eqref{eq:signr} preserves momentum but does not conserve energy. As such, our continuum description requires a dissipative mechanism to remove the energy injected by activity. This dissipative contribution can be included whilst preserving momentum by incorporating a bending viscosity. This viscosity is given, by analogy with the elastic bending moment and in agreement with other treatments in the literature~\cite{kodio_thesis,Buckmaster_Nachman_Ting_1975,salbreuxMechanics2017}, by
\begin{equation}
\boldsymbol{\Phi}^{\text{visc}}=\eta \dot\kappa {\bf t} {\bf b}\text{,}
\end{equation}
where $\eta$ is the bending viscosity. This viscous moment is associated with a stress tensor
\begin{equation}
\boldsymbol{\sigma}^{\text{visc}} = \eta \partial_s \dot{\kappa} {\bf t} {\bf n} \text{,}
\label{eq:sigvisc}
\end{equation}
which gives the following viscous force
\begin{equation}
    \bf {f} = \nabla \cdot {\boldsymbol \sigma}^\text{visc} = \eta (\partial_s^2 \dot{\kappa}{\bf n} + \kappa \partial_s \dot{\kappa}{\bf t})\text{.}
\end{equation}

In summary, the equations ${\boldsymbol \sigma}^{\mathrm{el}}$~\eqref{eq:sigel}, ${\boldsymbol \sigma}^{\mathrm{nr}}$~\eqref{eq:signr} and ${\boldsymbol \sigma}^{\mathrm{visc}}$~\eqref{eq:sigvisc} describe the mechanics of our nonreciprocal filaments---we now complete our equations of motion~\eqref{eq:ForceBalance2} with kinematic expressions for filament velocity $\dot{\bf {x}}$ and acceleration $\ddot{\bf {x}}$.

\subsection{Filament dynamics and key dimensionless parameters}
We build the dynamical equations for our active filaments by incorporating the total stresses ${\boldsymbol \sigma}={\boldsymbol \sigma}^\text{el}+ {\boldsymbol \sigma}^\text{visc}+{\boldsymbol \sigma}^\text{nr}$ into the force balance~\eqref{eq:ForceBalance2}. The resulting balance reads
\begin{align}
    &\rho \ddot{\bf x} + \beta \dot{\bf x} =  \partial_s \Lambda {\bf t} - \kappa \Lambda {\bf n} + A \left(\partial_s^2 \kappa + \frac{1}{2}\kappa\left(\kappa^2-\kappa_0^2\right)\right){\bf n}+  \zeta \left(\partial_s^{3} \kappa {\bf n} +  \kappa \partial_s^{2} \kappa {\bf t}\right)  + \eta \left(\partial_s^2 \dot{\kappa}{\bf n} + \kappa \partial_s \dot{\kappa}{\bf t}\right)\text{,}
    \label{eq:fullSystem}
\end{align}
supplemented with the incompressibility condition $\nabla \cdot {\bf \dot x} = 0$, and the geometric constraints ${\bf t} = \partial_s {\bf x}$, ${\bf n} = {\bf J }\cdot {\bf t}$ with ${\bf J}$ a right-handed rotation, and $ {\bf t}\cdot\partial_s {\bf n} = \kappa$. 

The system~\eqref{eq:fullSystem} is a nonlinear partial differential equation in the filament curvature, $\kappa(s,t)$, line tension $\Lambda(s,t)$, velocity $\dot{\bf x}(s,t)$ and acceleration $\ddot{\bf x}(s,t)$ parameterized by activity ${\zeta}$, substrate friction ${\beta}$, inertia $\rho$, viscosity $\eta$ and spontaneous curvature $\kappa_0$. For planar, inextensible curves the spontaneous curvature $\kappa_0$ simply renormalizes the line tension \cite{goldstein1995}, and only enters the equations via the boundary conditions of force and torque balance. For this reason we set spontaneous curvature to zero in the following analysis. 

This system has two natural non dimensionalisations.

\textit{Rescaling by viscous length and time scales---}The first natural non-dimensionalisation uses purely passive time and length scales,
\begin{equation}
    T= \frac{\eta}{A}, \quad L^2 = \frac{\eta}{\sqrt{\rho A}},
\end{equation}
giving a dimensionless nonreciprocity, friction and line tension
\begin{align}
   \bar\zeta =  \frac{\zeta}{A}\left( \frac{\sqrt{\rho A}}{\eta} \right)^{1/2}, \quad  \bar\beta =  \frac{\beta \eta}{\rho A}, \quad \bar\Lambda  = \frac{\Lambda\eta}{A\sqrt{\rho A}} \text{.}
   \label{eq:DimensionlessScheme1}
\end{align}
The viscous time scale $T$ describes the rate at which curvature dissipates via bending viscosity. The associated length scale $L$ gives a characteristic crossover length between viscous dissipation and inertial propagation. For wavenumbers $q>L^{-1}$, viscous dissipation will damp waves. For wavenumbers $q<L^{-1}$, flexural waves will propagate without dissipation. This scheme leaves a clear dimensionless nonreciprocity, $\bar\zeta$, to vary. However, to explore the overdamped limit $\rho\rightarrow 0$ this non-dimensionalisation is singular.

\textit{Rescaling by active length and time scales---} To explore the overdamped limit alternative length and time scale are provided by
\begin{equation}
    T= \frac{\eta}{A}, \quad L = \frac{\zeta}{A},
\end{equation}
with a dimensionless line density, friction and line tension given by
\begin{align}
    \epsilon = \frac{\rho \zeta^4}{A^2\eta^2}, \quad \quad \tilde\beta = \frac{\beta \zeta^4}{ A^4\eta},  \quad \quad \tilde\Lambda = \frac{\Lambda\zeta^2}{A^3} \text{.}
    \label{eq:DimensionlessScheme2}
\end{align}
This scheme has the disadvantage that the dependence on nonreciprocity is hidden within the natural lengthscale of the system. However, there is a clear overdamped limit $\epsilon\rightarrow 0$.  

In summary, we have introduced the nonlinear PDEs~\eqref{eq:fullSystem}, supplemented by geometric compatibility conditions, to capture the large deformations of nonreciprocal filaments. What is the resulting phenomenology?

\section{Tunable Nonreciprocal Instabilities}
\begin{figure*}[t!]
    \centering
    \includegraphics[width=\textwidth]{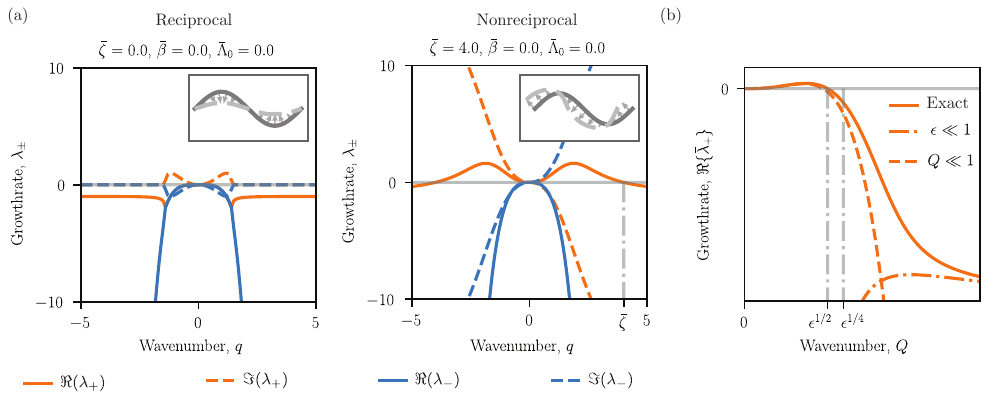}
    \caption{\textbf{Inertial instabilities of nonreciprocal filaments.} Linear stability analysis for the case of a close to flat filament with nonreciprocal stresses. The colour labels the two eigenvalues, $\lambda_+$ in orange and $\lambda_-$ in blue, with solid lines corresponding to real parts and dashed lines to the imaginary parts of the two eigenvalues. (a) Growth rates for a reciprocal filament, $\bar\zeta=0$, are always negative (stable). This includes decaying elastic inertial waves at small wavenumber. In the case of non-zero nonreciprocity, $\bar\zeta\neq 0$, the real part of eigenvalue $\lambda_+$ becomes positive with a non-zero imaginary part picking the advective direction for unstable waves to propagate. (b) The singular role of inertia is seen by taking overdamped limit $\rho\rightarrow 0 $. Here we nondimensionalize the dispersion~\eqref{eq:FlatFilament} instead using the active lengthscale $L=\zeta/A$~\eqref{eq:DimensionlessScheme2}, which exposes a dimensionless inertia $\epsilon=\rho \zeta^4/A^2\eta^2$ which we send to zero. A low inertia expansion of the dispersion~\eqref{eq:dimensionless2Polynomial} fails to resolve the instability as it breaks down for $Q<\epsilon^{1/4}$. Instead, we require a long wavelength expansion $Q\ll 1$ to resolve the instability. We also show where the real part of the eigenvalue changes sign at $Q=\epsilon^{1/2}$ or equivalently $q=\bar\zeta$ in the nondimensionalization used in panel (a).}
    \label{fig:InertialInstability}
\end{figure*}

To classify the instabilities of our nonreciprocal filaments, we first linearize~\eqref{eq:fullSystem} about the flat state parameterised by ${\bf x} = (x(s,t),h(s,t))$ where $h(s,t)\ll 1$. Our zeroth-order solution reads $\ddot{\bf x}=\dot{\bf x}=\kappa=h=0$, $\Lambda=\Lambda_0$: a static flat filament under a constant tension $\Lambda_0$. In this limit, the normal and tangential components of~\eqref{eq:fullSystem} decouple, and the normal component reads 
\begin{equation}
    \ddot{h} + \bar\beta \dot{h} =  \bar{\Lambda}_0 \partial_x^2h - \partial_x^4h - \bar\zeta \partial_x^{5} h - \partial_x^4\dot{h}.
    \label{eq:FlatFilament}
\end{equation}
Equation~\eqref{eq:FlatFilament} is a dynamic beam equation including friction, bending viscosity, and the crucial nonreciprocal coupling term $\bar{\zeta}\partial_x^5 h$. This term explicitly breaks the left-right symmetry of the beam via a polar bending moment $\boldsymbol{\Phi}^\mathrm{nr}=-\bar\zeta  h_{xxx} {\bf e}_x{\bf e}_z$, and will boost flexural waves unidirectionally.

Fourier transforming in space, $h(x,t)=\int\frac{\mathrm{d}q}{2\pi}\bar{h}(q,t)$, and looking for exponential growth in time of the form $\bar{h}(q,t)= \bar{h}(q) e^{\lambda t}$ we find the dispersion relation
\begin{equation}
    \lambda^2 + (\bar\beta +q^4)\lambda +q^4 + \bar\Lambda_0 q^2 + i \bar\zeta q^5 =0,
\end{equation}
giving growth rates of
\begin{equation}
    \lambda_\pm = \frac{1}{2} \left(-\bar\beta -q^4\pm\sqrt{(q^4+\bar\beta)^2-4 (q^4 + \bar\Lambda_0 q^2 + \mathrm{i} \bar\zeta  q^5)}\right).
    \label{eq:Roots}
\end{equation}

The dispersion~\eqref{eq:Roots} contains two distinctive nonreciprocal phenomena: an oscillatory instability driven by inertia (Fig.~\ref{fig:InertialInstability}) and the advection of otherwise stationary patterns (Fig.~\ref{fig:PrestressInstability}). 
\subsection{Inertial instabilities}
\label{subsec:inertial}
To exhibit the inertial instability, we set friction $\bar{\beta}=0$ and line tension $\bar{\Lambda}=0$ in~\eqref{eq:Roots} (Fig.~\ref{fig:InertialInstability}). In the absence of activity  the dispersion relation has negative growth rates, $\lambda_\pm$, with wavenumbers $|q|\lesssim 1$ advected in both directions along the filament with attenuation [Fig.~\ref{fig:InertialInstability}(a), reciprocal]. This is simply the passive viscoelastic relaxation of perturbations to a flat rod. As the nonreciprocity $\bar\zeta$ is switched on this left/right symmetry in advection is broken with one growth rate, $\Re(\lambda_+)$, becoming positive, whilst the other rate $\Re({\lambda_-})$ remains negative. Each of these modes is associated with waves traveling in opposite directions along the filament [Fig.~\ref{fig:InertialInstability}(a), nonreciprocal]. As the nonreciprocity is increased the activity actuates the higher curvature bending modes and the peak of $\Re(\lambda_+)$ progresses to higher wavenumbers. 

The wavevenumber for which the unstable mode stabilizes can be found exactly from~\eqref{eq:Roots} as
\begin{equation}
    \Re(\lambda_+)=0\implies q=\pm\bar\zeta \text{.}
\end{equation}
This band of unstable wavenumbers $q\in[-\bar{\zeta}, \bar{\zeta}]$ is inertial in the sense that it vanishes in the limit $\rho\rightarrow 0$ for fixed $\zeta$, because the dimensionless activity scales as $\bar{\zeta} \sim \zeta \rho^{1/4}$~\eqref{eq:DimensionlessScheme1}. To explicitly take the overdamped limit we switch to the dimensionless scheme~\eqref{eq:DimensionlessScheme2} to give the following reparameterization of the dispersion relation:
\begin{equation}\label{eq:dimensionless2Polynomial}
    \epsilon \tilde\lambda^2 +  Q^4\tilde\lambda + Q^4 + i Q^5=0\text{,}
\end{equation}
where $\epsilon= \rho \zeta^4/A^2\eta^2$ is the dimensionless inertia and $Q$ is the dimensionless wavenumber. In these units the band of unstable modes is given by
\begin{equation}
  \Re(\tilde\lambda_+)=0\implies Q=\epsilon^{1/2}\text{,}  
  \label{eq:Qroot}
\end{equation}
which vanishes as $\epsilon \rightarrow 0$.

The dispersion~\eqref{eq:dimensionless2Polynomial} can be expanded in powers of $\epsilon$ in the low inertia regime under the assumption that $ \epsilon^{1/4} \ll Q $ and $\epsilon\ll 1$ to give
\begin{align}
    &\tilde\lambda_+ = (-1-i Q) +\frac{(Q-i)^2 \epsilon }{Q^4}+ O\left(\epsilon ^2\right)\text{,}\label{eq:lambdaPLowreynolds}\\
    &\tilde\lambda_-=-\frac{Q^4}{\epsilon }+(1+i Q)-\frac{(Q-i)^2 \epsilon }{Q^4}+O\left(\epsilon ^2\right)\text{,}\label{eq:lambdaMLowreynolds}
\end{align}
up to quadratic order in $\epsilon$. 
The second root, $\tilde{\lambda}_{-}$, in~\eqref{eq:lambdaMLowreynolds} is a fast-decaying mode that becomes singular as $\epsilon\to 0$. The first root, $\tilde\lambda_+$~\eqref{eq:lambdaPLowreynolds} gives the overdamped limit---the leading term in $\tilde\lambda_+$ captures elastic decay and pure advection. We can see the inertial nature of the instability by noting that for $\epsilon\ll 1$ and $\epsilon^{1/4}\ll Q$, this low $\epsilon$ expansion of $\tilde\lambda_+$ can never capture the real zero of the full polynomial~\eqref{eq:Qroot} because $\epsilon^{1/4}>\epsilon^{1/2}$ [Fig.~\ref{fig:InertialInstability}(b), dot-dashed line].
 
The instability can instead be captured by considering the inertial regime and expanding for $Q\ll \epsilon^{1/4}$ as follows:
\begin{align}
    &\tilde\lambda_+ = -\frac{i Q^2}{\sqrt{\epsilon }}+\frac{Q^3}{2 \sqrt{\epsilon }}-\frac{Q^4 \left(4+i \sqrt{\epsilon }\right)}{8 \epsilon }-\frac{Q^5}{16 \sqrt{\epsilon }}+O\left(Q^6\right)\text{,}\\
    &\tilde\lambda_- =\frac{i Q^2}{\sqrt{\epsilon }}-\frac{Q^3}{2 \sqrt{\epsilon }}+\frac{i Q^4 \left(\sqrt{\epsilon }+4 i\right)}{8 \epsilon }+\frac{Q^5}{16 \sqrt{\epsilon }}+O\left(Q^6\right)\text{.}
\end{align}
The real part of this small $Q$ expansion of $\tilde\lambda_+$ is shown in Fig.~\ref{fig:InertialInstability}(b) along with the exact and low inertia expansions. These two expansions show that the instability is indeed inertial and that it is separate from the underlying advection of curvature by the bending moments, a phenomena which is preserved even in the zero inertia regime, see the first term in~\eqref{eq:lambdaPLowreynolds}. 

\subsection{Pre-stress driven instabilities}
\begin{figure*}[t]
    \centering
    \includegraphics[width=\textwidth]{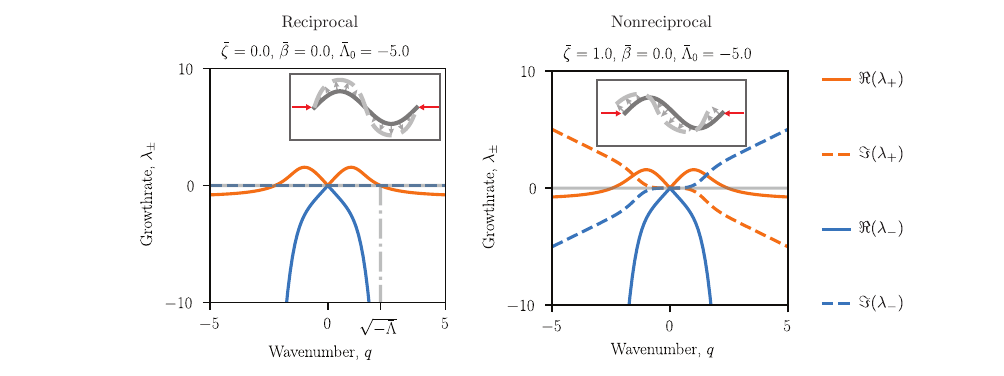}
    \caption{\textbf{Prestressed reciprocal and nonreciprocal instabilities of filaments.} Linear stability analysis for a close to flat filament with nonreciprocal stresses and prestress (negative line tension, $\bar\Lambda_0<0$). The colour labels the two eigenvalues, $\lambda_+$ in orange and $\lambda_-$ in blue, with solid lines corresponding to real parts and dashed lines to the imaginary parts of the two eigenvalues. (Left) Reciprocal case shows a classical buckling instability, with a band of unstable wavenumbers $\Re(\lambda)>0$ between $q=\pm\bar \Lambda_0$. The eigenvalues have no imaginary part and thus there are no flexural waves on the filament. (Right) By contrast, when nonreciprocity is coupled to this prestress there is again a buckling instability, but now with associated imaginary part that changes sign at $q=0$, thus giving one way flexural waves in the curvature generated by the prestress.}
    \label{fig:PrestressInstability}
\end{figure*}
The zero inertia limit gives rise to an interesting observation: that curvature advection can essentially be decoupled from the inertial active instability, that is to say, even in the absence of inertial forces the nonreciprocal terms will advect curvature. To explore this scenario further we consider the case where the beam is under prestress, that is a negative tension $\bar\Lambda_0<0$ (Fig.~\ref{fig:PrestressInstability}). When $\bar{\zeta}=0$ we find a classical buckling instability in~\eqref{eq:Roots} with characteristic lengthscale
\begin{equation}
    q=\pm \sqrt{-\bar\Lambda_0},
\end{equation}
and no wave propagation, $\Im(\lambda)=0$ (Fig.~\ref{fig:PrestressInstability}, left). When nonreciprocity is turned on the curvature generated by this buckling is advected, leading to nonzero imaginary parts to the dispersion relation [Fig.~\ref{fig:PrestressInstability}, right]. Note that only one direction of wave motion is associated to a growing mode---at the nonlinear level, we expect unidirectional pulsations of curvature.

\subsection{Breaking momentum conservation with frictional interactions}
\begin{figure*}[t]
    \centering
    \includegraphics[width=\textwidth]{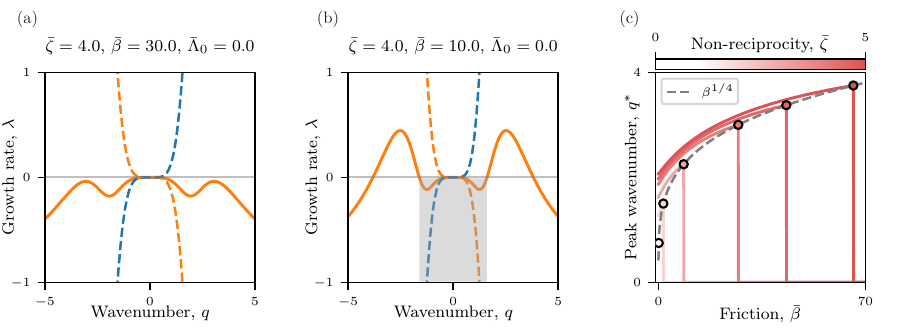}
    \caption{\textbf{Frictional gapping of the inertial instability.} (a,b) Linear stability analysis for a close to flat filament with nonreciprocal stresses and frictional interaction with the substrate ($\bar\beta>0$). The colour labels the two Eigenvalues, $\lambda_+$ in orange and $\lambda_-$ in blue, with solid lines correspond to real parts and dashed lines to the imaginary parts of the two Eigenvalues. As friction decreases [$\beta=30$ in panel (a), $\beta=10$ in panel (b)] we find a band of unstable wavenumbers opens about a finite $q$ which is gapped from the origin--- The highlighted grey region around $q=0$ in (b) shows where the frictional interactions have stabilized the filament at long wavelengths. (c) Plot of the peak of the real part of the growth rate, $\text{argmax}_q \lambda_\pm$, against dimensionless friction, $\bar{\beta}$, for a range of nonreciprocal activities, $\bar\zeta$, alongside our prediction for a threshold friction $q^*=(3\bar\beta)^{1/4}$. Increased friction screens long wavelength modes leading to a shift in the dominant lengthscale of instability to shorter lengths before the instability is completely screened by friction.}
    \label{fig:frictionScreening}
\end{figure*}
The instabilities described above occur whilst conserving momentum---what is the effect of breaking this conservation law via dissipative interactions? Focusing on the inertial instability described in \S\ref{subsec:inertial}, we now include a finite environmental friction $\bar{\beta}$ in the dispersion~\eqref{eq:Roots}. We find that friction gaps the range of unstable wavenumbers and introduces a threshold activity for instability, turning a long-wavelength instability into finite wavenumber patterns [Fig.~\ref{fig:frictionScreening}(a,b)]. Demanding that both $\Re(\lambda)=0$ and a repeated root in~\eqref{eq:Roots} gives a criteria for the dominant wavenumber of the instability $q^*$ in terms of a critical activity $\zeta^*$ and friction $\beta^*$~[Fig.~\ref{fig:frictionScreening}(c)]:
\begin{equation}
q^*=\frac{3}{4} \bar\zeta^* = (3 \bar\beta^*)^{1/4}.
\end{equation}

In conclusion, our stability analysis gives a series of predictions for advective instabilities driven either by inertia (Fig.~\ref{fig:InertialInstability}) or pre-stress (Fig.~\ref{fig:PrestressInstability}), with a dominant wavenumber tuneable via environmental interactions (Fig.~\ref{fig:frictionScreening}). However the fate of these instabilities beyond the linear regime remains unclear---to address this question, we now examine their nonlinear evolution by solving the full dynamics~\eqref{eq:fullSystem} numerically.

\section{Nonlinear dynamics and pattern formation in a nonreciprocal ring}
\begin{figure*}[t]
    \centering
    \includegraphics[width=\textwidth]{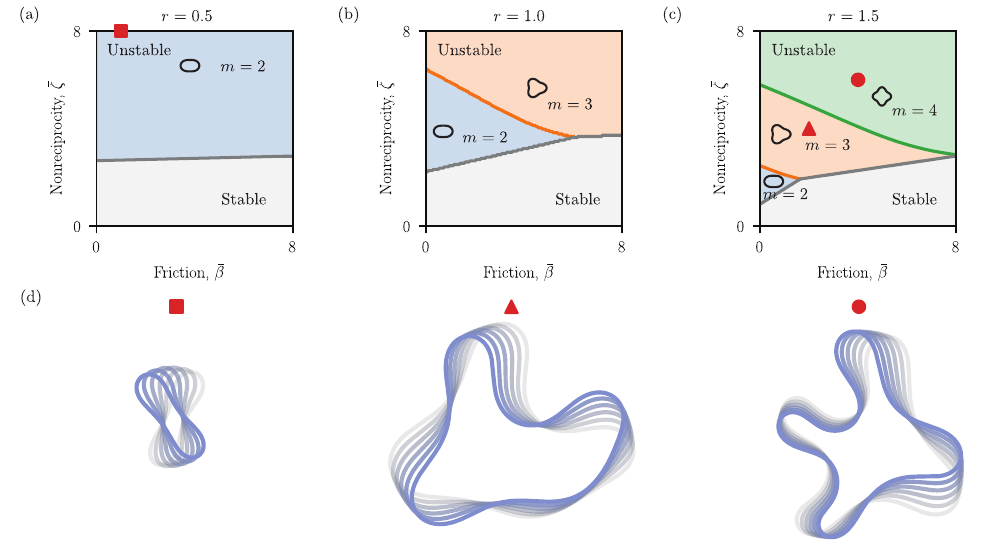}
    \caption{\textbf{Inertial instability of an odd ring} Linear stability phase portraits showing the most unstable mode for a ring of radius $r$ with substrate friction, $\bar\beta$, plotted against nonreciprocity, $\bar\zeta$. The gray area denotes that stable region, blue correspond to an $m=2$ instability, orange to $m=3$ and green to $m=4$. We plot this for radii of $r=0.5,1,1.5$ (a), (b) \& (c). (b) \& (c) demonstrate how frictional gapping can be used to excite higher order modes before stabilising by taking a line of constant nonreciprocity, $\bar\zeta$, and varying increasing substrate friction, $\bar\beta$. In panel (d) we show numerical solutions to the full equations in the nonlinear regime demonstrating the first three modes of instability for parameters shown in (a) \& (c) with transparency indicating previous timepoints.}
    \label{fig:ringLinStab}
\end{figure*}

To study the nonlinear evolution of our filaments, we focus on a ring---a symmetric geometry with simple boundary conditions that is often used to study elastic pattern formation~\cite{box2020,kodio2020}. We integrate the system~\eqref{eq:fullSystem} numerically by discretizing curvatures and velocities along the filament and time-stepping using an implicit scheme with modified Powell solver---our full numerical scheme is described in Appendices \S\ref{subsec:Kinematics} and \S\ref{sec:appNumerics}. To 
seed our simulations, we use modes obtained via a linear stability analysis on the ring, which gives the dispersion [Appendix~\S\ref{sec:RingStability}]
\begin{align}
   \left(m-m^3\right)^2 (r+i \bar\zeta  m)+\bar\beta  \lambda  \left(m^2+1\right) r^5+\lambda ^2 \left(m^2+1\right) r^5+\lambda  m^2 \left(m^2-1\right)^2 r=0\text{,}
   \label{eq:RingStability}
\end{align}
for the $m^\mathrm{th}$ Fourier mode of a ring of radius $r$. This dispersion predicts the same qualitative features as the flat state [Fig.~\ref{fig:ringLinStab}(a)]: at zero friction, modes $m=2, \cdots, {\bar\zeta}^{1/2} r$ become unstable,  with finite friction $\bar\beta$ increasing both the threshold activity and dominant wavenumber of the instability [Fig.~\ref{fig:ringLinStab}(b,c)]. The nonlinear behavior of these modes is shown in Fig.~\ref{fig:ringLinStab}(d) with examples of the nonlinear growth of the first $3$ $m$ modes as predicted by the linear stability analysis. 
\begin{figure*}
    \includegraphics[width=\textwidth]{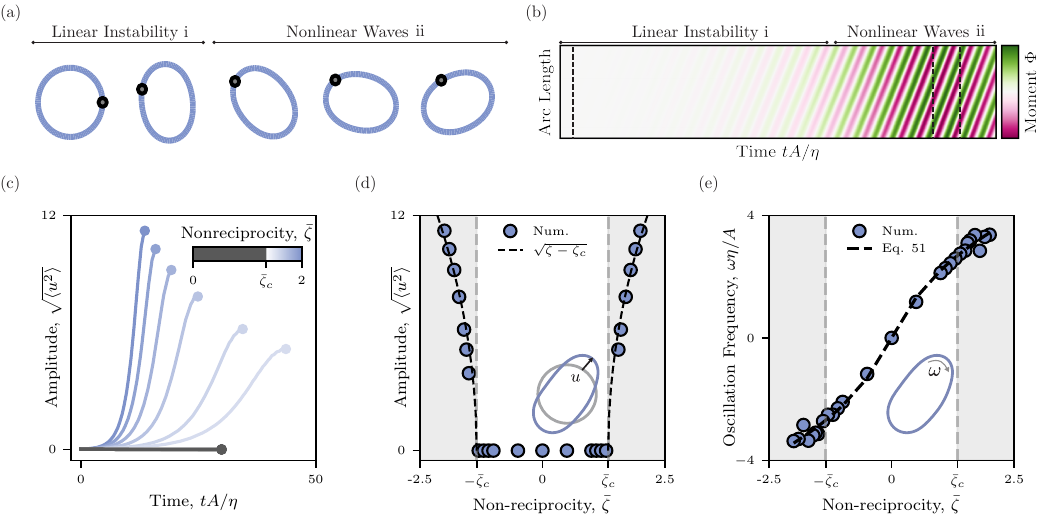}
    \caption{\textbf{Curvature advection in the nonlinear regime.} Beyond linear instability, a closed ring exhibits stable nonlinear traveling patterns. (a,b) Snapshots of cyclic ring dynamics, and kymographs showing the emergence of travelling torque waves. A single material point along the ring is highlighted---because motion is driven by cyclic shape change, not rigid rotation, this material point appears to propagate opposite the direction of wave travel. Snapshot series (i) and (ii) correspond to regions (i) and (ii) of the kymograph. (c-e) Nonlinear waves emerge via a supercritical Hopf bifurcation. (c,d) The amplitude of deformations $\sqrt{\langle u^2 \rangle}$, with $u$ the displacement field, as a function of time and activity. We find this amplitude grows as $\sqrt{\bar \zeta - \bar \zeta_c}$ beyond a threshold activity $\zeta_c\approx1.34$ captured by the linear dispersion Eq.~\ref{eq:RingStability}. (e) Cycles emerge with a finite frequency beyond $\zeta_c$, with a chirality tied to the sign of $\zeta_c$, prior to this threshold the frequency is that of the damped oscillations. The linearized frequencies predicted by Eq.~\ref{eq:RingStability} also capture this weakly nonlinear regime.}
    \label{fig:ringNonlinear}
\end{figure*}

Driving the ring beyond its linear inertial instability, we find flexural waves which are stabilised by the bending elastic geometric nonlinearities [Fig.~\ref{fig:ringNonlinear}(a,b)]. Focusing on activities $\bar \zeta$ at which the lowest $m=2$ mode is linearly unstable [Fig.~\ref{fig:ringLinStab}(a)], at the nonlinear level we find a single traveling pattern occupies the entire ring, causing a persistent chiral cycling motion. Systematically varying the activity $\bar \zeta$ reveals that this pattern appears via a supercritical Hopf bifurcation [Fig.~\ref{fig:ringNonlinear}(c-e)]. Oscillation amplitudes grow continuously beyond a threshold activity $\bar \zeta_c \approx 1.34 \dots$ as $\sqrt{\bar \zeta - \zeta_c}$  [Fig.~\ref{fig:ringNonlinear}(d)], and emerge with a finite frequency which then increases with activity [Fig.~\ref{fig:ringNonlinear}(e)]. This threshold $\bar \zeta_c$ is captured by our stability analysis~\eqref{eq:RingStability}, and we find that our linearized predictions for cycling frequency remain accurate even into the nonlinear regime.

\begin{figure*}[t]
    \includegraphics[width=\textwidth]{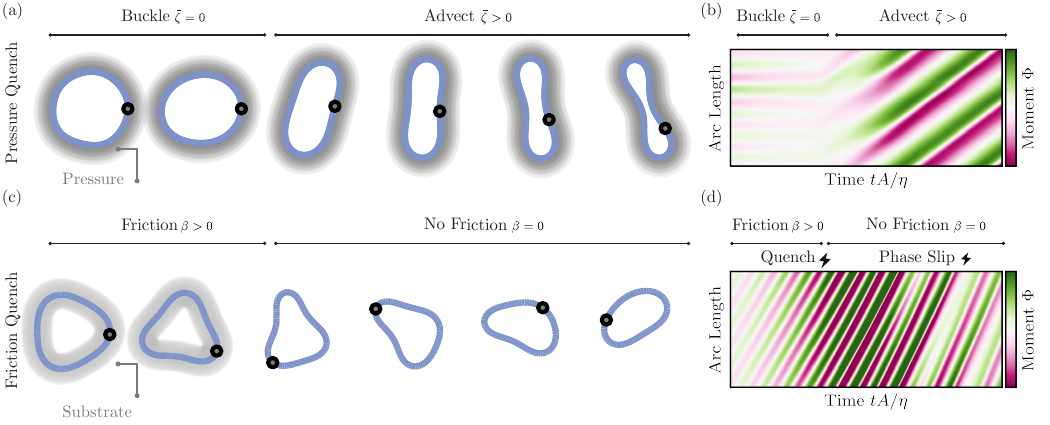}
    \caption{\textbf{Tuneable nonlinear patterns from the environment.}
    (a-b) A surface tension drop, or external pressure, will buckle the filament, generating curvature which is then advected by nonreciprocity. Under a  pressure drop $\Delta \gamma=\chi (A_0^{-1}-A^{-1})$, with $\chi=10$ and $A_0=0.3\pi r^2$, we apply a step change in activity, from $\bar \zeta=0$ to $\bar \zeta=1$. The ring transitions from a static buckled shape into travelling nonlinear waves.
    (c-d) A friction quench will cause the ring to hop between modes via a series of phase slips. Here we quench from $\beta=15$ to $\beta =0$ at $\bar\zeta=4$, driving an $m=3$ to $m=2$ transition predicted by our stability analysis (Fig.~\ref{fig:ringLinStab}).}
    \label{fig:ringNonlinearExternal}
\end{figure*}

Our linear buckling instability also persists into the nonlinear regime. Setting $\zeta=0$, we first buckle the ring using a surface tension difference $\Delta \gamma$ between the ring interior and exterior:
\begin{equation}
    \Delta\gamma = \chi\left(\frac{1}{A_0}-\frac{1}{A}\right)\text{,}
\end{equation}
with $\chi$ an inverse compressibility, $A$ the current ring area and $A_0$ a target area, in a similar manner to a collapsing elastica-bounded soap film [Fig.~\ref{fig:ringNonlinearExternal}(a)]~\cite{box2020,kodio2020}. We then apply a step-change in activity $\bar\zeta$, and find that the buckled geometry is immediately advected by nonreciprocity, resulting in persistent cycles in shape space [Fig.~\ref{fig:ringNonlinearExternal}(b)], this transition to advected buckled patterns is an exceptional transition, a hallmark of nonreciprocal systems \cite{fruchart2021,fruchart2023,al-izzi2025}. Buckling induced limit cycles have recently been observed in clamped follower-force chains~\cite{zheng2023}, and nonreciprocal robotic metamaterials~\cite{al-izzi2025}, where active waves preferentially pile up at one end of the filament to drive snapping. In clamped beams, these waves must overcome the energy barrier of elastic snap-through to drive oscillations. By contrast, in symmetric buckling every buckled state of the ring is degenerate---nonreciprocity can actuate this family of modes without any barrier.

Finally, we show that even post-instability the selected pattern can be tuned by environmental interactions. In Fig.~\ref{fig:ringNonlinearExternal}(c-d) we quench the filament between a high friction and low friction environment. At large friction $\beta$, high wavenumber patterns---here $m=3$---dominate [Fig.~\ref{fig:ringLinStab}(b)]. Upon a quench to $\beta=0$ however, we observe a series of phase slips [Fig.~\ref{fig:ringLinStab}(d)] down to the lowest $m=2$ mode predicted by our stability analysis~\eqref{eq:RingStability}. In conclusion, the nonlinear dynamics is extremely robust---we can tuneably hop between stable patterns with inertia, pre-stress or dissipation.

\section{Discussion}
In this work we have developed a geometrically nonlinear continuum theory for nonreciprocal active filaments, and discovered both inertial and pre-stress induced instabilities that seed persistent nonlinear waves. In contrast to other geometrically nonlinear models of active filaments, such as follower force active filaments \cite{decanio2017,stein2021,man2020,clarke2024,zheng2023,chakrabarti2019,sangani2020} or mechano-chemical systems~\cite{cassReactiondiffusionBasisAnimated2023}, these waves break time reversal symmetry whilst remaining net force-and-torque free, without requiring additional internal variables.

Seen from a design perspective, such waves offer a compelling toolkit for engineering soft robotic locomotion. Living locomotors self-organize flexural waves for crawling~\cite{brackenburyCaterpillarKinematics1997} or rolling~\cite{liangMechanicsSoftBodyRolling2025} gaits. By contrast, synthetic mimics rely on centralized control to drive oscillations~\cite{dengLightsteerableLocomotionUsing2024,dengNonlinearWavesFlexible2021, kornerNonlinearBeamModel2020} and precisely sculpt their timing~\cite{huSmallscaleSoftbodiedRobot2018}. Breaking reciprocity offers a particularly simple route to self-organized travelling patterns, which has recently been exploited to design crawling, walking and digging in flexible mechatronic linkages~\cite{al-izzi2025, veenstra2025}. Our continuum formulation emphasizes that such travelling patterns are generic, and could be functionalized not only in mechatronics but across physical platforms where reciprocity breaking has been observed, for example colloidal chains~\cite{chajwaWavesAlgebraicGrowth2020} or microfluidic devices~\cite{beatusPhononsOnedimensionalMicrofluidic2006}.

Our theory is also an example of inertia bringing new instabilities, nonlinear dynamics, and functionality to active matter. The canonical examples of active matter---solid or fluidic systems internally driven by energy-non-conserving processes---are overdamped. However many artificial and natural active systems such as robotic swarms~\cite{scholzRotatingRobotsMove2018}, starling flocks or inertial active Brownian particles~\cite{lowenInertialEffectsSelfpropelled2020} are inherently massive. The introduction of an inertial timescale, with associated wave propagation, can radically alter the phenomenology of these systems relative to their overdamped limit. In our active filaments inertia gives a boundary layer effect and a corresponding instability which cannot be found in the purely viscous regime~\cite{chatterjeeInertiaDrivesFlocking2021}. Whilst inertial effects can lead to chaos and turbulence~\cite{sarkarMechanochemicalFeedbackDrives2025a}, here the instabilities that we uncover instead lead to well-controlled rhythmic motion. We anticipate further exciting mechanical phenomena at the intersection of wave physics and active matter.

There are many possible extensions to the model proposed here. Development of rigorous, convergent symplectic schemes which accurately conserve length and momentum could open the door to more efficient and longer simulations \cite{bergou_discrete_2008,marsden_multisymplectic_1998,nemeth_shape-space_2026} along with developing robust simulations based on using tools from discrete differential geometry on the variational forms of this problem~\cite{tongDiscreteDifferentialGeometry2026,gazzola_forward_2018}. Such methods are often harder to prove formal convergence to the corresponding continuous problem, but are much better behaved numerically \cite{alougesNlink2025}, and thus very useful for exploring a wide variety of scenarios including environmental interactions.

On a theoretical front allowing the filament to deform in $3$D rather than being confined to a plane would allow for a much richer phasespace of wave motion.  The phenomonology in this case would become more complex and additional nonreciprocal couplings that couple twist and bend are possible. Allowing additional degrees of freedom in the constitutive relations, \textit{e.g.}~shearablity of the filament, will also open up the possibility of other odd moduli, such as those discussed in Ref.~\cite{chen2021,nemeth2025}. The interplay between multiple nonreciprocal stresses in the geometrically non-linear regime and more complex environmental interactions (\textit{e.g.}~those driven by gravity), could enable the engineering of complex cycles of shape change, and be a valuable design too for developing autonomous, bioinspired, soft robotics.

\appendix
\section{Filament kinematics and intrinsic equations of motion}
\label{subsec:Kinematics}
To solve the PDE~\eqref{eq:fullSystem} numerically, we first decompose the dynamics into the normal and tangential velocities of the filament itself. Our filament moves with velocity 
\begin{equation}
\dot{\bf x} = W{\bf t} + U {\bf n}\text{,}
\label{eq:vel}
\end{equation}
with $W$ the tangential, and $U$ the normal velocity. The condition of inextensibility is given by taking the divergence of the velocity field and using the definition of the curvature
\begin{equation}
    \nabla \cdot \dot{\bf x}=\partial_s W + U \kappa = 0\text{,}
    \label{eq:Incompressibility}
\end{equation}
where $\nabla = {\bf t}\partial_s$ is the curve divergence operator.

In order to describe the acceleration of the rod $\ddot{{\bf x}}$ we must calculate time derivatives of the tangent $\bf t$, normal $\bf n$ and curvature $\kappa$. Taking the time derivative of our tangent vector gives
\begin{align}
    &\partial_t {\bf t} =  \partial_t \partial_s {\bf x} = \partial_s\partial_t {\bf x} = \partial_s\left( W {\bf t} + U {\bf n} \right), \nonumber\\
    &= \left( \partial_s W + U \kappa\right) {\bf t} + \left( \partial_s U - W \kappa\right) {\bf n}, \nonumber\\
    &= \left( \partial_s U - W \kappa\right) {\bf n},
\end{align}
where the last step has used the incompressibility condition~\eqref{eq:Incompressibility}.
Similarly, for the change in normal vector and curvature:
\begin{align}
    &\partial_t {\bf t} = \left( \partial_s U - W \kappa\right) {\bf n} \notag \\
    &\partial_t {\bf n} = \left( W \kappa - \partial_s U\right) {\bf t} \notag \\
    &\partial_t \kappa =  \partial_t ( - {\bf n}\cdot\partial_s {\bf t})= \partial_s (\kappa W - \partial_s U).
    \label{eq:timederivatives}
\end{align}
Finally, we have an expression for the acceleration $\ddot{\bf{x}}$ in the induced frame of the filament by taking the time derivative of~\eqref{eq:vel} and using~\eqref{eq:timederivatives}:
\begin{equation}
\ddot{\bf x} = \left[\dot{W} + U(\kappa W - \partial_s U)\right]{\bf t}
+ \left[\dot{U} + W(\partial_s U - W \kappa)\right]{\bf n}\text{.}
\label{eq:acceleration}
\end{equation}

supplemented with the incompressibility condition $\partial_s W + \kappa U = 0$, and the geometric constraints: ${\bf t} = \partial_s {\bf x}$, ${\bf n} = {\bf J }\cdot {\bf t}$, and $ {\bf t}\cdot\partial_s {\bf n} = \kappa$. Rewriting the force balance~\eqref{eq:fullSystem} using the the kinematic expressions for filament velocity $U, W$~\eqref{eq:vel} and~\eqref{eq:acceleration} we find 
\begin{align}
\label{eq:fullSystem1}
&\rho \left[  \dot{W} + U(W \kappa - \partial_s U) \right] + \beta W = \partial_s \Lambda +  \zeta \kappa \partial_s^{2} \kappa +\eta \kappa \partial_s \dot{\kappa} \, \mathrm{Tangential\ Balance}  \\
&\label{eq:fullSystem2} \rho \left[  \dot{U} + W(\partial_s U - W \kappa) \right] + \beta U = - \Lambda  \kappa + A \left( \partial_s^2 \kappa + \frac{1}{2}\kappa\left(\kappa^2-\kappa_0^2\right)^2\right) + \zeta \partial^3_s\kappa+ \eta \partial_s^2 \dot\kappa  \, \mathrm{Normal\ Balance}\\
&\label{eq:fullSystem3}\partial_s W + \kappa U =0, \quad \mathrm{Continuity} \\
&\label{eq:fullSystem4}\dot{\bf x} = W{\bf t} + U{\bf n}\text{,} \quad \mathrm{Filament\ Position} \\
&{\label{eq:fullSystem5}\bf t} = \partial_s {\bf x}\text{,} \quad {\bf n} = {\bf J }\cdot {\bf t}\text{,}\quad {\bf t}\cdot\partial_s {\bf n} = \kappa \quad \mathrm{Geometrical\ Relations} \text{.}
\end{align}

In dimensionless scheme (\ref{eq:DimensionlessScheme1}) the full set of coupled equations for our active filaments are given by
\begin{empheq}[box=\fbox]{align}\label{eq:NDSystem1}
&\textbf{Dimensionless Scheme} \notag \\
 &\dot{W} + U(W \kappa - \partial_s U)  + \bar\beta W = \partial_s \bar\Lambda +  \bar\zeta \kappa \partial_s^{2} \kappa + \kappa \partial_s^2(W \kappa - \partial_s U), \quad \mathrm{Tangential\ Balance}  \\
  &\label{eq:NDSystem2}\dot{U} + W(\partial_s U - W \kappa) + \bar\beta U = -  \bar\Lambda  \kappa
 +  \partial_s^2 \kappa + \frac{1}{2}\kappa\left(\kappa^2-\kappa_0^2\right) + \bar\zeta \partial^3_s\kappa+  \partial_s^3 (W \kappa - \partial_s U), \quad \mathrm{Normal\ Balance}\\
&\label{eq:NDSystem3}\partial_s W + \kappa U =0, \quad \mathrm{Continuity} \\
&\label{eq:NDSystem4}\dot{\bf x} = W{\bf t} + U{\bf n}\text{,} \quad \mathrm{Filament\ Position} \\
&{\label{eq:NDSystem5}\bf t} = \partial_s {\bf x}\text{,} \quad {\bf n} = {\bf J }\cdot {\bf t}\text{,}\quad {\bf t}\cdot\partial_s {\bf n} = \kappa \quad \mathrm{Geometrical\ Relations} \text{,}
\end{empheq}
where we have made use of the intrinsic expression for the curvature rate, Eq.~\ref{eq:timederivatives}, to rewrite the viscous forces in terms of the filament velocities $W$ and $U$. It is these dimensionless equations that we will solve numerically.

\section{Numerical scheme}\label{sec:appNumerics}
Here we take a practical approach to solving these PDEs numerically where we discretize up to second order based on approximations that assume variations in the mesh spacing are small. We split the time derivatives using a backwards Euler method as
\begin{equation}
    \dot X_i \approx \frac{X^{t}_i-X^{t-1}_i}{\Delta t} 
\end{equation}
and spatial derivatives using a central difference style second order accurate stencil
\begin{align}
    &\partial_s X^t_i \approx D_s X^t_i= \frac{X^{t}_{i+1}-X^{t}_{i-1}}{2\Delta s^{t}_i}\\
    & \partial^2_s X^t_i \approx D^2_s X^t_i= \frac{X^{t}_{i+1}-2 X^{t}_{i}+X^{t}_{i-1}}{(\Delta s^{t}_i)^2}
\end{align}
where $\Delta s^{t}_i$ is the discrete Hodge dual to the discrete distance $1$-form on the line, that is
\begin{equation}
    \Delta s^t_i = \frac{1}{2} \left( |{\bf x}^{t}_{i+1}-{\bf x}^{t}_{i}| + |{\bf x}^{t}_{i}-{\bf x}^{t}_{i-1}| \right)
\end{equation}
with $|\cdot|$ denoting the $\mathrm{L}^2$ norm.

For the dynamical equations this gives
\begin{align}
         & {\bf t}^{t+1}_i = \frac{D_s {\bf x}^{t+1}_i}{|D_s {\bf x}^{t+1}_i|}\\ 
         & {\bf x}^{t+1}_i = {\bf x}^{t}_i + \Delta t \left[ W^{t+1}_i {\bf t}^{t+1}_i + U^{t+1}_i {\bf J}\cdot{\bf t}^{t+1}_i \right]\\
         & \kappa^{t+1}_i = - ({\bf J}\cdot{\bf t}^{t+1}_i)\cdot D_s {\bf t}^{t+1}_i\\
         &D_s W^{t+1}_i + \kappa^{t+1}_i U^{t+1}_i = 0\\
         &  W^{t+1}_i - W^{t}_i + \Delta t U^{t+1}_i(W^{t+1}_i \kappa^{t+1}_i - D_s U^{t+1}_i) + \bar\beta \Delta t W^{t+1}_i = \Delta t \big(D_s \Lambda^{t+1}_i + \bar\zeta \kappa_i^{t+1}(\nabla^2\kappa)^{t+1}_i\nonumber\\
         &\quad +  \kappa_i^{t+1} \left(D^2_s(\kappa_i^{t+1}W_i^{t+1}-D_s U_i^{t+1}\right)\big) \\
         &  U^{t+1}_i -  U^{t}_i + \Delta t W^{t+1}_i(D_s U^{t+1}_i - W^{t+1}_i \kappa^{t+1}_i) + \bar\beta \Delta t U^{t+1}_i =  - \Delta t  \kappa^{t+1}_i  \Lambda^{t+1}_i+ \bar\zeta \Delta t D_s(\nabla^2\kappa)^{t+1}_i\nonumber\\
         &\quad+ \Delta t ((\nabla^2\kappa)^{t+1}_i + \frac{1}{2} (\kappa^{t+1}_i)^3) +  \Delta t D_s^3 (\kappa^{t+1}_i W^{t+1}_i - D_s U^{t+1}_i) \\
         &(\nabla^2\kappa)^{t+1}_i = D_s^2\kappa^{t+1}_i\text{.}
\end{align}

We solve this non-linear system using a modified Powell rootfinder found in the Hybrid scheme in SciPy \cite{scipy}. The proof of stability for such a scheme is an open problem, however we find that it agrees well for the relaxation of a passive ring although we encounter some issues with length loss during longer simulations (a common problem with such simplistic numerical schemes).

\section{Stability of a ring}
\label{sec:RingStability}
We consider a ring of base radius, $r$, and a position and time dependent defection from this radius $u(s,t)$. We will assume that this can be parameterised by the single-valued function $u(\theta,t)$ where $\theta$ is the usual polar angle. We can write this curve parameterised by the vector
\begin{equation}
    {\bf x} = r[ 1 + u(\theta,t)]\left( \cos \theta,\sin\theta\right),
\end{equation}
and arclength derivative given by $\partial_s = r^{-1}[1+u]^{-1} \partial_\theta$. 

We will proceed to analyse everything in this section up to linear perturbations in the defection from the base radius, $u$. The tangent and normal vectors to first order in perturbation $u$ are given by
\begin{align}
    & {\bf t} = \left( -\sin \theta + u_\theta \cos\theta,\cos\theta + u_\theta \sin\theta \right)\\
    & {\bf n} = \left(-\cos\theta -u_\theta \sin\theta, -\sin \theta + u_\theta \cos\theta \right)
\end{align}
and curvature by
\begin{equation}
    \kappa = -\frac{1}{r} + \frac{u+u_{\theta\theta}}{r}\text{,}
\end{equation}
where the first term denotes the curvature of the base ring and the second the corrections due to the deformation.

The filament velocity in the normal direction is given by $U = - r \dot u$ and we write the tension as $\Lambda = \Lambda_0 + \delta\Lambda$ where $\delta\Lambda\sim u$ is the variation in tension along the filament needed to impose inextensibility. We also assume that the filament velocity is small, $U,W\sim u$, and then linearise the equations of motion in order to find dynamical equations for the deformation $u$.

Tangential and normal force balance components are given by
\begin{align}
    &\dot{W}+\bar\beta W=-\frac{\bar\zeta  \left(u_{\theta\theta}+u_{\theta\theta\theta\theta}\right)}{r^4}-\frac{\dot{u}_{\theta} + \dot{u}_{\theta\theta\theta}}{r^3}+\frac{\delta \Lambda_\theta }{r},\label{eq:ringTangentialForce}\\
    &-r(\ddot{u}+\beta   \dot{u})=\frac{\bar\zeta  \left(u_{\theta\theta\theta} + u_{\theta\theta\theta\theta\theta}\right)}{r^4}+\frac{u_{\theta\theta} + u_{\theta\theta\theta\theta}}{r^3}+\frac{\dot{u}_{\theta\theta} +\dot{u}_{\theta\theta\theta\theta}}{r^3}+\frac{1}{2} \left(\frac{r^2\kappa_0^2-1}{r^3}-\frac{\left(\kappa_0^2 r^2-3\right) \left(u_{\theta\theta} + u\right)}{r^3}\right)\nonumber\\
    &\qquad+\frac{-\Lambda_0 \left(u_{\theta\theta} + u\right)+\delta \Lambda +\Lambda_0}{r}\text{,}\label{eq:ringNormalForce}
\end{align}
where the $\theta$ subscripts denote differentiation and we have made use of the inextensibility condition
\begin{equation}
    W_\theta = - r\dot u\text{,}
\end{equation}
to eliminate the tangential velocity $W$ from the normal force balance equation.

At zeroth order in $u$ we find the following condition on the ground state line tension needed to maintain a ring of constant radius $r$ (and spontaneous curvature $\kappa_0$) in the unperturbed state
\begin{equation}
    \Lambda_0 = \frac{1}{2}  \frac{1}{r^2}- \frac{1}{2}\kappa_0^2\text{,}
\end{equation}
which we will take as our value of $\Lambda_0$ from here on.

We now turn our attention to the higher order force balance conditions. Taking a derivative in $\theta$ of Eq.~\eq{eq:ringTangentialForce} and discrete Fourier transforming in the angle $\theta$, $f=\sum_m\bar{f}_m e^{im\theta}$ with $m\in\mathbb{Z}$, and, assuming each mode has time dependance of the form $\bar{u}_m=\bar{u}_m(0)e^{\lambda t}$, such that $\lambda$ is growth rate of the $m$th mode, gives an expression for the variation in line tension needed to keep the filament inextensible,
\begin{equation}
    \delta\bar{\Lambda}_m = \bar{u}_m\frac{-\lambda  m^4 r +\lambda  m^2 r +i \bar{\zeta}m^3\left(  1-  m^2 \right)+\bar\beta  \lambda  r^5 +\lambda^2 r^5}{m^2 r^3}\text{.}
\end{equation}
This expression can then be substituted into the discrete Fourier transformed version of the force balance in the normal direction, Eq.~\eq{eq:ringNormalForce}, and we arrive at the following dispersion relation
\begin{align}
\left(m-m^3\right)^2 (r+i \bar\zeta  m)+\bar\beta  \lambda  \left(m^2+1\right) r^5+\lambda ^2 \left(m^2+1\right) r^5+\lambda  m^2 \left(m^2-1\right)^2 r=0\text{.}
\end{align}

This equation can be solved for the growth rates, $\lambda_\pm(m,\bar\zeta,\bar\beta,r)$, to find the fastest growing $m$-mode for a set of chosen parameters. We plot the mode rate with maximal growth rate, $\mathrm{argmax}_m \{\Re(\lambda_\pm)\}$, as a function of friction, $\bar\beta$, and nonreciprocity, $\bar\zeta$, in Fig.~\ref{fig:ringLinStab}, for several values of radius, $r$.

\section*{Acknowledgments}
This research includes computations using the computational cluster Katana supported by Research Technology Services at UNSW Sydney. S.C.A. was supported by funding from the European Union’s Horizon Europe research and innovation programme under the Marie Skłodowska-Curie postdoctoral fellowship No. 101106384, the Australian Research Council Centre of Excellence for the Mathematical Modelling of Cellular Systems (MACSYS, CE230100001), and an Australian Research Council Discovery Early Career Researcher Award (DECRA, DE260101426). J.B. was supported by funding from the European Union’s Horizon Europe research and innovation programme under the Marie Skłodowska-Curie postdoctoral fellowship No. 101106500. The authors thank R.G.~Morris, A.~Souslov, J.~Veenstra, S.J.A.~Poulain \& G.P.~Alexander for helpful discussions.
\bibliography{bibliography}

\end{document}